# Novel Superconductivity in Endohedral Gallide $Mo_8Ga_{41}$

P.Neha,[1†] P.Sivaprakash,[2†] K.Ishigaki,[3] G.Kalaiselvan,[2] K.Manikandan,[2] Y.Uwatoko,[3] R.S. Dhaka,[4] S. Arumugam,[2*] S. Patnaik[1*]

[1]School of Physical Sciences, Jawaharlal Nehru University, New Delhi, India-110067

[2]Centre for High Pressure Research, School of Physics, Bharathidasan University,

Tiruchirappalli, Tamilnadu - 620024, India

[3]Institute for Solid State Physics, University of Tokyo, Kashiwa, Chiba 277-8581, Japan

[4]Department of Physics, Indian Institute of Technology, Hauz Khaz, New Delhi 110016, India.

†These Authors contributed equally

*Corresponding author: spatnaik@mail.jnu.ac.in, sarumugam1963@yahoo.com

**Abstract:** We report on synthesis and characterization of gallide cluster based $Mo_8Ga_{41}$ superconductor. Transport and magnetization measurements confirm the superconducting transition temperature to be 9.8 K. The upper critical field ($H_{c2}$), lower critical field ($H_{c1}$), Ginzburg-Landau coherence length ($\xi_{GL}$) and penetration depth($\lambda$) are estimated to be 11.8T, 150G, 5.2nm, 148nm respectively. The electronic band structure, density of states and phonon dispersion curve calculations are obtained by using Density Functional Theory. The core level X-ray Photoelectron Spectroscopy (XPS) reveals the binding energy information of the constituting elements Mo and Ga in $Mo_8Ga_{41}$. The valence band spectra from XPS is in good agreement with calculated density of states (DOS). The zero field critical current density ($J_c$) at T = 2 K is ~ $3 \times 10^5$ A/cm$^2$ which is indicative of efficient flux pinning in the as grown compound. About two fold enhancement in critical current density with application of external pressure (1.1 GPa) is observed with marginal decrease in transition temperature. The fitting of current density to double exponential model confirms possibility of two gap superconductivity in $Mo_8Ga_{41}$.

**Introduction**

Intermetallic compounds are the original warehouse of novel superconductivity [1-2]. In the recent past, the gallium based intermetallic clusters have attracted considerable attention towards attaining new superconducting phases [3-10]. In this family, typically a transition element is surrounded by gallium atoms and these have emerged as a favorable structural motif for superconductivity. Several endohedral gallide cluster based superconductors such as $Mo_8Ga_{41}$, $Ir_2Ga_9$, $PuCoGa_5$, $ReGa_5$ [6-10] have been reported with transition temperature as high as 18 K [7]. In particular, the Mo based gallide clusters are relatively easy to make but their microstructure optimization for high current carrying applications is pending. With regard to microscopic origin of superconductivity in $Mo_8Ga_{41}$, a strong electron phonon coupling mechanism has been established [9] and recent muon spectroscopy data confirm the presence of two-gap superconductivity. Appearance of superconductivity is also reported by doping V at Mo site, along with emergence of charge density wave (CDW) correlation [11].

From the atomic coordination point of view, molybdenum atoms in $Mo_8Ga_{41}$ are trapped inside gallium cages and each Mo atom forms ten coordination bonds with surrounding Ga. This cluster of Ga cages, with transition metals at the center, plays the key role for the occurrence of superconductivity in endohedral gallide cluster based binary and ternary intermetallic compounds [8]. The composition of binary gallium superconducting system is simply represented by $Mo_xGa_{5x+1}$ and for x = 6, 8 the superconducting transition temperature for $Mo_6Ga_{31}$, $Mo_8Ga_{41}$ is reported to be 8 K, 9.8 K respectively [9][10]. Moreover, application of superconducting materials rests on their optimal higher upper critical field ($H_{c2}$) and critical current density ($J_c$). In general the current density is enhanced by intrinsic defects such as secondary precipitates or cold-worked dislocations or by extrinsic defects created by ion irradiation induced columnar or point defects that leads to increase in pinning centers [12][13]. The effectiveness of flux pinning and consequent enhancement in current density of a superconductor relate to both pinning center density and magnitude of pinning force strength that crucially depends on optimized microstructure. In this paper we report synthesis and characterization of polycrystalline $Mo_8Ga_{41}$ superconductor. Further, using Density Function Theory (DFT)[15-17] we obtain the electronic band structure of the synthesized compound. We also report X-ray photoelectron spectroscopy(XPS) studies which yields the core level binding energy of constituent elements of the compound. The effect of pressure on superconducting transition temperature ($T_c$) and critical current density ($J_c$) are also estimated.

**Experiment**

The polycrystalline samples of $Mo_8Ga_{41}$ were synthesized using solid state reaction by taking constituent elements Mo (99.999%) powder and Ga (99.999%) pieces in proper stoichiometry in a quartz ampoule and by evacuating the tube down to $10^{-4}$ mbar. The evacuated tube was heated at 850 $^0$C for 55 hour and very slowly cooled down to 170 $^0$C. Finally shiny grey colored samples were obtained. Room-temperature X-ray diffraction measurement was done on the powdered samples by the RIGAKU powder X-ray Diffractometer (Miniflex-600) with Cu-K$\alpha$ radiation. Scanning Electron Microscopy (SEM) images were obtained from *Bruker AXS Microanalyser* and from a *Zeiss EVO40* SEM analyzer respectively. High-resolution photoelectron spectroscopy measurements have been performed using a commercial Omicron ESCA+ system equipped with hemispherical electron energy analyzer and a monochromatic Al x-ray source (hν = 1486.6 eV) from Oxford Instruments, Germany. The overall energy resolution was better than 0.6 eV using 50 eV analyzer pass energy and monochromatic x-ray source. The vacuum level in the analysis chamber during the measurements was better than $5\times10^{-10}$ mbar.

Since the XPS technique is very surface sensitive, the samples were cleaned *in-situ* by light sputtering (using $Ar^+$) to remove the contamination from the sample surface. The Mo 3d and Ga 2p core-level spectra have been fitted by using a Voigt function including the instrumental broadening.

The transport measurements were performed using linear four probe technique using copper wire and silver epoxy. Magnetization measurements were carried out using 14T Cryogenic Physical Property Measurement System (PPMS)**.** At various pressures magnetization measurements were performed using Magnetic Property Measurement System (MPMS, Quantum Design, USA). The nonmagnetic Cu–Be alloy were used to fabricate the clamp type miniature hydrostatic pressure cell to generate external pressure upto 1GPa. For pressure measurements, the fluorinert FC#70 and FC#77 (1:1) mixture was used as pressure transmitting medium and the in-situ pressure (P) was estimated by considering the superconducting transition of pure Sn as a manometer. Temperature dependence of magnetization M(T) was recorded upon zero field cooling at various pressures under external magnetic field of 20 Oe . For resistivity measurement under pressure up to 2GPa, the HPC-33 Piston type pressure cell was used in Physical Property Measurements System (PPMS-14T, Quantum Design). Hydrostatic pressures was applied using BeCu/NiCrAl clamped piston- cylinder cell immersed in Fluorinert (FC70:FC77=1:1) pressure transmitting medium. A cubic anvil device was used for electrical resistivity measurements for various pressures from 1.5 to 8 GPa. The sample was immersed in a pressure medium of Daphne #7373 oil to maintain the hydrostatic pressure and encapsulated in a Teflon cell, surrounded by a pyrophyllite block. This block was evenly compressed from six directions using six tungsten carbide (WC) anvils. The six WC anvils crush the pyrophyllite gasket and compress the Teflon cell from six directions equally and the hydrostatic nature of the pressure is maintained beyond the solidification of the pressure medium. Furthermore, the pressure is controlled and kept constant during warming and cooling cycle of the cubic press device.

**Results and Discussion**

The main panel of Figure 1(a) shows the Rietveld refined XRD powder data of polycrystalline $Mo_8Ga_{41}$ using FULLPROF software. The XRD peaks reveal that the material crystallizes in $V_8Ga_{41}$ type crystal structure with R-3(148) space group. All the peaks could be satisfactorily indexed, confirming phase purity of the sample. The lattice parameters are estimated to be a = b = 14.04 Å, c = 15.05 Å, density 7.1g/cm$^3$ which are in agreement with previously reported values [9]. The inset (i) in Figure 1(a) shows the scanning electron microscopy (SEM) image of prepared compound which clearly indicates the existence of grains with varying grain size. Figure 1(b) shows the schematic crystal structure of $Mo_8Ga_{41}$ where each Mo atom forms ten coordination bonds with surrounding Ga atoms that create the endohedral cluster. In such structural motif, Mo atoms occupy two crystallographic positions and Ga occupies nine crystallographic positions [8, 9].

**Band structure calculations**

Figure 2 deals with the theoretically calculated electronic band structure, density of states (DOS) and phonon dispersion curve of $Mo_8Ga_{41}$ by employing density function theory (DFT) using Vienna Ab-initio Simulation Package (VASP 5.4)[15]. For theoretical calculation, projected augmented wave (PAW) [16] with plane wave basis set of cut off energy 282.691eV is used. The DFT calculations were performed using GGA-PBE (Generalized Gradient Approximation- Perdew Burke Ernzerhof) approximation to define exchange correlation energy

[17]. To achieve the optimal convergence, typically 9×9×9 mesh point with 0.074×0.074×0.074 K-point grid was employed. The unit cell used for theoretical calculation comprises of three formula units. The band structure of $Mo_8Ga_{41}$ is shown in Figure 2(a) which presents the dispersion of bands in all high symmetry direction. Crossing of bands between conduction and valence band through $E_F$ illustrates the metallic nature of material which is in agreement with the experimental results. Narrow dispersion of bands supports very strong correlation in the material which is reflected in transport measurement results as well. The theoretically calculated density of states (DOS) is shown in Figure 2(b). The DOS around Fermi energy is mainly contributed by Mo 4d and Ga 4p orbital hybridization. Here total density of states (DOS) around $E_F$ is around 30eV [in formula unit] which indicates metallic behavior and superconductivity in the material. In lower energy side, the highest peak comes from Mo 3d and Ga 2p orbital contribution. In endohedral gallide based binary intermetallic superconductors, the $E_F$ is found to lie in pseudo gap which is unlike usual superconductors in which the $E_F$ occupies a state with finite DOS. The placement of $E_F$ in a pseudo gap is attributed to the splitting of the degenerate orbitals of Ga at $E_F$ in the presence of transition elements forming the endohedral gallide structure. The position of the $E_F$ in the pseudo gap provides the chemical stability to the compounds [8].

The phonon dispersion curve is shown in Figure 2(c). Here, very large number of branches in longitudinal optical (LO), longitudinal acoustical (LA), transverse optical (TO) and transverse acoustical (TA) branches are attributed to the presence of large number of atoms in the unit cell taken for theoretical calculation which contains three formula unit. The gap in the dispersion curve can be ascribed to the mass difference of the Mo and Ga atom. The phonon dispersion curve is showing only positive frequencies which rule out the presence of structural instability in the compound. Recently, it is documented that on V doping at Mo place in

Mo$_8$Ga$_{41}$, CDW phase is observed[11]. Evidently, for the parent compound, the phonon dispersion curve comprise only positive frequencies that rules out the possibility of CDW type lattice instability theoretically.

**X-ray Photoelectron Spectroscopy**

The X-ray Photoelectron spectroscopy (XPS) study for Mo$_8$Ga$_{41}$ is shown in Figure 3. The core level binding energy spectra of Mo$_8$Ga$_{41}$ corresponding to the constituent elements Mo and Ga are shown in Figure 3(a) and (b). The core band spectra corresponding to the Mo 3d state (Figure 3(a)) where two strong peaks corresponding to 3d$_{3/2}$ and 3d$_{5/2}$ states are localized at binding energy 230.4 eV and 227.2 eV respectively. The second components corresponding to 3d$_{3/2}$ and 3d$_{5/2}$ states are localized at binding energy 231.4 eV and 228.2 eV respectively. The obtained experimental results are in close agreement with other core level XPS studies made on Mo based compounds [18]. Further, for the Ga 2p state the core band spectra spilt into two peaks at binding energy 1143.4 eV and 1116.5 eV for 2p$_{1/2}$ and 2p$_{3/2}$ states respectively. The second components for the Ga 2p state the core band spectra spilt into two peaks at binding energy 1144.4 eV and 1117.8 eV for 2p$_{1/2}$ and 2p$_{3/2}$ states respectively (shown in Figure 3(b)). There are some satellite peaks observed at 1131.7 eV. Figure 3(c) presents the valence band (VB) spectra for Mo$_8$Ga$_{41}$ and Figure 3(d) shows the theoretically calculated DOS for total Mo and Ga. The density and occupancy of the electronic states in valence band can be easily illustrated from the VB spectra. The comparison of the VB spectra with respect to theoretically calculated DOS indicates very good agreement between the two. The valance band spectra shows the feature at about the energies around -2.3eV, -5eV, -9 eV. Very sharp feature of about -2.3eV is very well matched with theoretically calculated total DOS and it is contributed mainly by Mo d and Ga p orbitals mixing. The feature at around -5eV, -9 eV occurs due to contribution of Mo d and Ga s, p orbitals.

**Transport measurements**

The variation of electrical resistivity of $Mo_8Ga_{41}$ with respect to temperature is plotted in Figure 4(a). The onset of superconducting transition is seen at 9.8 K while zero resistivity state is achieved at 9.4 K (inset). In the normal state, and above 100 K, the change in resistivity grows marginally. The magneto-resistance measurements data are shown in inset (i) of Figure 4(b). The superconducting transition temperature shifts to lower temperature gradually with applied external field up to 10 Tesla. The broadening in the transition with applied magnetic field is attributed to the formation of vortex state. The variation of transition temperature with applied magnetic field yields the H-T phase diagram, which is shown in inset (ii) of Figure 4(b). The main panel of Figure 4(b) shows the extrapolation of H-T phase diagram fitted with Ginzburg Landau formula $H_{c2}(T)=H_{c2}(0)[(1-(T/T_c)^{1/2})/(1+(T/T_c)^{1/2})]$. This fitting yields the upper critical field $H_{c2}(0) = 11.8$ T. The Ginzburg–Landau coherence length is estimated (by using the formula $\xi_{GL}= (\Phi_0/2\pi H_{c2})^{1/2}$, where $\Phi_0=2.07\times10^{-7}$ G cm$^2$) to be $\xi_{GL}(0) \sim 5.28$nm. In $Mo_8Ga_{41}$, existence of surface superconductivity is documented where by superconductivity could be sustained above $H_{c2}$ ($H_{c3}=1.69H_{c2}$)[9]. The calculated $H_{c2}$ from the transport measurements remains higher than magnetization or specific heat. The nucleation of superconductivity on surface becomes more pronounced when the applied magnetic field is parallel to the sample. In such case, the vortex lattice overlaps cause the nucleation of the superconductivity to the surface [19]. Further, for type II superconductors the pair breaking is caused either by spin paramagnetic or orbital effect. Under orbital limiting case, pair breaking takes place due to enhancement of kinetic energy of pair compared to condensate state in presence of magnetic field. In Pauli paramagnetic limit the spin alignment with applied magnetic field breaks the Cooper pairs under favourable energy condition. In clean limit, the orbital upper critical field is given by $H_{c2}^{orbital}(0) = -0.72T_c[dH/dT]_{T=T_c}$ which is estimated as 7.76T. The $H_{c2}^{pauli}$ on the other hand is given by 1.86 $T_c$ that equals to 18.2 Tesla. Corresponding to orbital and Pauli upper critical field, the Maki parameter is calculated as $\alpha=1.414 H_{c2}^{orbital}/H_{c2}^{pauli}=0.6 < 1$ which indicates that the superconductivity in the $Mo_8Ga_{41}$ is Pauli limited.

In magneto-resistance measurements, broadening in resistivity curve is ascribed to the vortex motion in mixed state. The Arrhenius plot, shown in main panel of Figure 4(c), provides flux flow activation energy dependence on applied magnetic field. The flux flow activation energy ($U_0$) variation with applied magnetic field is related by the relation

$$\rho = \rho_0 \exp(-U_0/K_B T) \qquad (1)$$

where $U_0$ is field dependent activation energy $\rho_0$ is field independent pre-exponential factor. The variation of dissipation of energy with magnetic field due to vortex motion in mixed state which in turn provides the scale for potential application is shown in inset of Figure 4(c). The straight line behavior in Arrhenius plot validates the thermally activated flux flow (TAFF) process as per Arrhenius law. The activation energy $U_0$ shows different power law dependence for different magnetic field range. The Arrhenius plot provides the activation energy dependence on field as $U_0(B) \alpha B^{-0.62}$ (for $B < 5T$) and $U_0(B) \alpha B^{-2.7}$ ($B > 5T$).

**Magnetization measurements**

Next we turn to the magnetization measurement performed under zero field cooling (ZFC) and field cooling (FC) protocols which is shown in Figure 5(a). The characteristic diamagnetic transition is obtained at 9.8 K. A broad irreversible region between ZFC and FC curves indicate good flux pinning in the sample. The inset of Figure 5(a) shows isothermal M-H curve taken at T = 2 K. From it, the lower critical field ($H_{c1}$) is estimated to be 150 G by noting the deviation from linear behavior in M-H loop in the lower field region. Corresponding to $H_{c1}$, the penetration depth ($\lambda$) is given by $H_{C1} = \varphi_0/(2\pi\lambda^2)$ which is estimated to be 148 nm. The critical current density ($J_c$ at 2K) is calculated using Bean's critical state model formula; $J_c = 20\Delta M/a[1-a/3b]$ where a(0.5mm), b(1.5mm) are thickness and width of the sample on which the measurements are made such that a < b and $\Delta M$ is the magnetization difference of the M-H curve when field is increasing (M+) and decreasing (M-) at a particular field. The sample was placed in PPMS such that the magnetic field direction is perpendicular to the cross-sectional area of width and thickness. The calculated current density is of the order of $3\times10^5$ A/cm$^2$. The variation of $J_c$ with respect to field is shown in Figure 5(b).

**High Pressure study**

The effect of external pressure on the superconducting properties of $Mo_8Ga_{41}$ is presented in Figure 6. Figure 6(a) shows variation of electrical resistivity as a function of applied external pressure till 6 GPa. We observe that on increasing pressure, the $T_c$ decreases drastically at 1.5 GPa but then it starts to increase from 8 K to 8.5 K as the pressure is increased upto 6 GPa. The suppression of transition temperature at low pressure is also confirmed in magnetization measurement that is shown in Figure 6(b). Variation of transition temperature with pressure is shown in inset of Figure 6(b). Above 1.5 GPa, the pressure coefficient, $dT_c/dP$ is estimated to be 0.07482 K/GPa. We note that in similar Ga cluster based superconductor $PuCoGa_5$, enhancement in transition temperature from 18 K to 22 K [7] is reported on application of external pressure. Such increase in $T_c$ with pressure reflects possibility for higher $T_c$ phases of Ga cluster based samples by fine tuning electron-phonon coupling parameter. From Figure 6 (a) we see that as the pressure increases the normal state resistivity of the material also decreases. Under BCS theory, $T_c \sim \Theta_D \exp(-1/N(E_F))V_0$ where $\Theta_D$ is the Debye temperature, $V_0$ is electron phonon coupling strength and $N(E_F)$ is related to $m^* n^{1/3}$ where $m^*$ is effective mass and n is carrier density. Here $T_c$ is increasing with pressure and metallicity is also increasing with pressure which in turn indicates increasing carrier concentration (n) with pressure. In general, intermetallic compounds are generally found to be very susceptible toward pressure variations because applied pressure can relatively tune lattice instability. As seen in Figure 4a, the slow growth of normal state resistivity in higher temperature range stays fairly unchanged with applied pressure and emergence of any kind of lattice instability is not indicated [16]. The magnetization measurements taken under ZFC-FC protocol in the presence of external pressure from 0GPa to 1.1 GPa are shown in figure 6(b). The blank circles and filled circles in the graph indicate the ZFC and FC data that are in agreement with low pressure resistivity data. From magnetization data we note that the transition temperature decreases from 9.8 K to 9.6 K under applied pressure of 1.1 GPa. The inset of Figure 6(b) shows the variation of $T_c$ with pressure from the combined resistivity and magnetization data.

Figure 7 presents the variation of critical current density $J_c$ as a function of external magnetic field, temperature and pressure that is derived from magnetization loops by using Bean's model formula. The sample was placed such that the magnetic field direction remained parallel to larger dimension of the specimen. Figure 7 shows the variation of critical current density with magnetic field taken at T = 5 K and applied pressure 0 GPa ($P_0$),0.2 GPa ($P_1$), 0.4 GPa ($P_2$), 0.72 GPa ($P_3$), and 1.1 GPa ($P_4$). About two fold enhancement of in critical current density ($J_c$) is achieved at 1.1GPa compared to ambient pressure ($J_c$). The relatively large $J_c$ is reflective of effectiveness of as grown defects in our polycrystalline samples for strong pinning. Thus there is clear evidence for increase in pinning force density with pressure that implies still higher $J_c$ can be achieved by optimizing defect morphology of polycrystalline $Mo_8Ga_{41}$. The enhancement in critical current density was also observed for data taken at 7 K and 9 K implying the impact of pressure on pinning center formation continues close to transition temperature. This is more striking because the transition temperature shows decreasing trend at low pressure and therefore the actual enhancement in $J_c$ would be more than what is observed [11,12]. The enhancement in $J_c$ with pressure indicates improved pinning density and grain connectivity with applied pressure.

**Two gap Superconductivity**

Recent muon spectroscopy measurement performed on $Mo_8Ga_{41}$ indicates the presence of conventional two-gap s-wave superconductivity [13]. The presence of two gap model can also be tested by critical current density variation with magnetic field by using double exponential model [14]. Figure 8 presents the variation of $J_c$ (in log scale), taken at 2K, 5K, 7 K and 9K with respect to the reduced field h (h=H/$H_{irr}$). Here $H_{irr}$ stands for irreversibility field that is experimentally determined as the crossover point between reversible and irreversible regions in the positive H axis of M-H curve. We note that $J_c$ is decreasing exponentially with respect to magnetic field. The fitting of the $J_c$ vs h curve is in good agreement with double exponential model

$$J_c(h) = J_1 \exp(-A_1 h) + J_2 \exp(-A_2 h) \qquad (2)$$

where, $J_1$ and $J_2$ are the partial critical current density corresponding to each gap and $A_1$, $A_2$ are constants. The good agreement of the data with double exponential model confirms the possibility for the presence of two gap superconductivity in the material. The variation of $J_1$, $J_2$ taken at different temperature is shown in inset of Figure 8. Small difference in $J_1$, $J_2$ magnitude indicates there is less difference in two energy gaps values and our data are in good agreement with the applicability of two gap s-wave model for $Mo_8Ga_{41}$.

**Conclusion**

In summary, we have synthesized the polycrystalline samples of $Mo_8Ga_{41}$ superconductor and determined the characteristic parameters like $T_c$, $H_{c1}$, $H_{c2}$, $\lambda$, $\xi$ as 9.8K, 150 G, 11.8 T, 5.28nm and 148 nm respectively through combined transport and magnetization measurements. Theoretical calculation of band structure, density of states and phonon dispersion curve is performed using Density Functional Theory (DFT). Only positive frequencies in the phonon dispersion curve rule out the existence of structural instability like charge density waves (CDW) which is in agreement with the experimental results. The binding energy information for the Mo(3d) and Ga(2p) state in $Mo_8Ga_{41}$ is obtained from the core level X-ray Photoelectron Spectroscopy (XPS). The pressure studies yield the suppression in the superconducting transition temperature in the lower pressure range (0-1.5 GPa) while a slight increase in $T_c$ is observed with positive pressure coefficient 0.07482 K/GPa on increasing the pressure (1.5-6GPa). The zero field critical current density($J_c$), calculated at 2K is of the order of $10^5$ A/cm$^2$ indicates good pinning in the material. Around two fold enhancement in zero field critical current density (5K) is observed with applied pressure of 1.1GPa although the transition temperature decreased. The double exponential model fitting confirms applicability of two gap model to $Mo_8Ga_{41}$.

**Acknowledgements:**

The author PN acknowledges University Grant Commission (UGC) for providing SRF under Basic Science Research (BSR) fellowship. SA acknowledges BRNS (Mumbai), DST (SERB, FIST, PURSE), CEFIPRA, DRDO, and UGC (SAP, PURSE, FIST), New Delhi. SP thanks the DST-FIST program and project EMR/2016/003998 of the Department of Science and Technology, Government of India for low temperature high magnetic field facilities at JNU. Technical support from AIRF (JNU) is acknowledged.

**Figure Captions:**

**Figure 1.** (a) Reitveld refined XRD data of $Mo_8Ga_{41}$. Scanning electron microscopy (SEM) image of $Mo_8Ga_{41}$ are shown in the inset. (b) Schematic crystal structure of $Mo_8Ga_{41}$.

**Figure 2.** (a) Band structure (b) Density of states (DOS) and (c) Phonon dispersion curve of $Mo_8Ga_{41}$ calculated by DFT.

**Figure 3.** X-ray Photoelectron Spectroscopy of $Mo_8Ga_{41}$ (a) electronic charge state of Mo (b) electronic charge state of Ga (c) Valance Band spectra of $Mo_8Ga_{41}$ (d) Comparison of theoretical DOS with VB spectra.

**Figure 4.** (a) Resistivity variation as a function of temperature. Superconducting transition as $T_c^{onset}$ and $T_c^{zero}$ is seen in inset (b) Extrapolated upper critical field $H_{c2}(0)$ is plotted with GL fitting. Inset (i) shows magnetoresistance measurements with external magnetic field 0T,1T,1.5T,2T,2.5T,3T,3.5T,4T,4.5T,5T,6T,7T,8T,9T,9.5T,10T; Inset (ii) shows H-T phase diagram (c)Arrhenius plot of resistvity for 0T,1T,1.5T,2T,2.5T,3T,3.5T,4T,4.5T,5T,6T,7T,8T,9T,9.5T,10T magnetic field. Inset shows activation energy dependence on applied field.

**Figure 5.** (a) Magnetization measurement under ZFC-FC protocol. M-H loop at T = 2 K is shown in the inset (b) Critical current density ($J_c$) variation with respect to external magnetic field.

**Figure 6.** (a) Resistivity variation as a function of temperature at different applied pressure is plotted. Onset transition temperature variation in magnified scale is seen in inset(i). (b) Magnetization measurement under ZFC-FC protocol for different applied pressure 0 GPa ($P_0$), 0.2 GPa ($P_1$),0.4 GPa ($P_2$),0.72 GPa ($P_3$),1.1 GPa ($P_4$) are included. The unfilled and filled circles are for magnetization data taken under FC and ZFC protocol respectively. Inset shows variation of transition temperature ($T_c$) with pressure from resistivity and magnetization data.

**Figure 7.** (a)Variation of critical current density ($J_c$)at different applied pressure for $P_0$, $P_1$, $P_2$, $P_3$ ,$P_4$ (0 GPa, 0.2GPa,0.4GPa,0.72GPa,1.1 GPa) respectively at T = 5 K. Inset shows variation of pinning force $F_p$ with pressure.

**Figure 8.** (a) Variation of critical current density ($J_c$) taken at 2 K, 5 K, 7 K and 9 K as a function of reduced field h that is fitted with double logarithmic model (b) Variation of partial critical current density corresponding to each gap ($J_1$ and $J_2$ )(from Equation 2) with temperature.

**Figure 1**

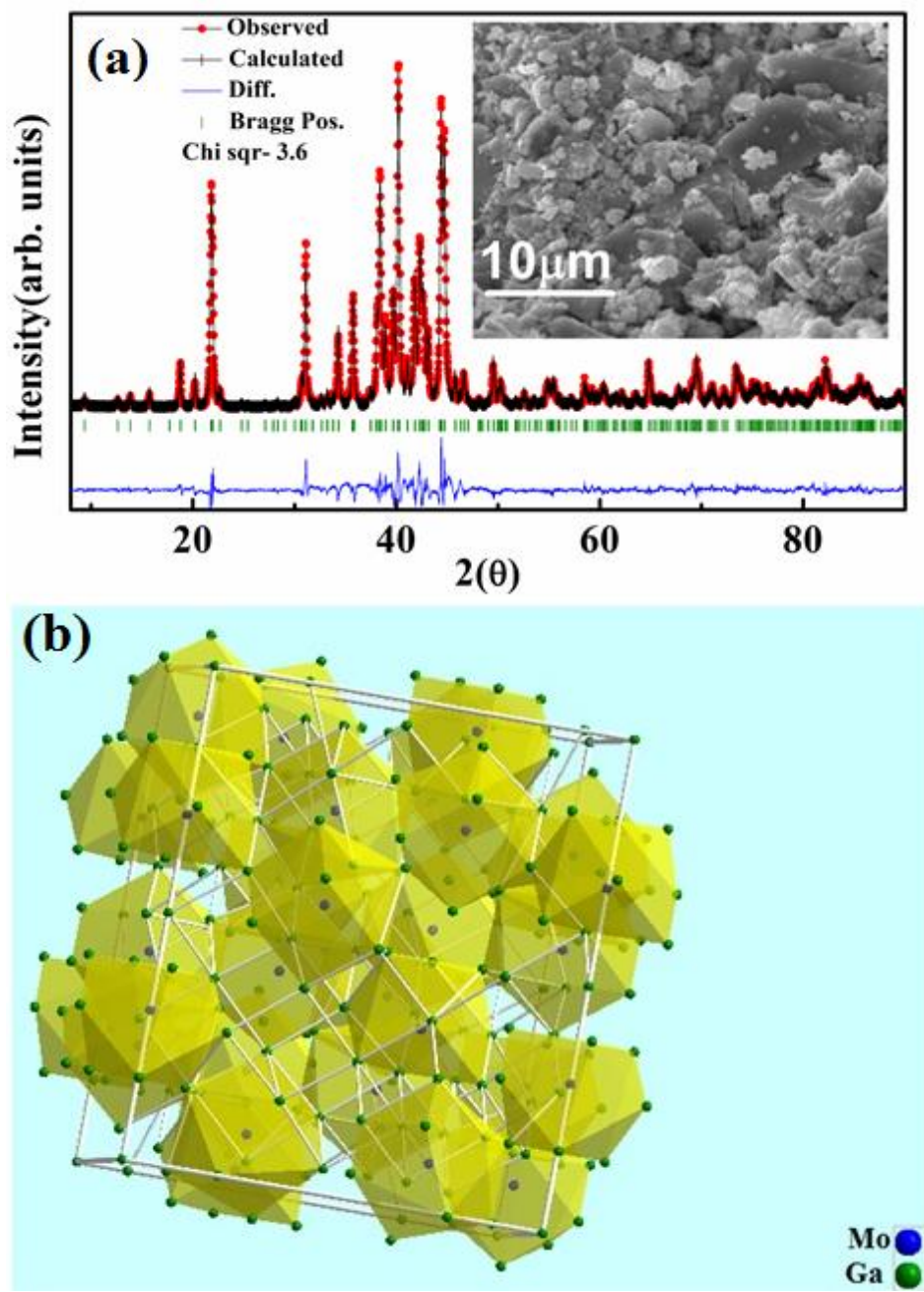

**Figure 2**

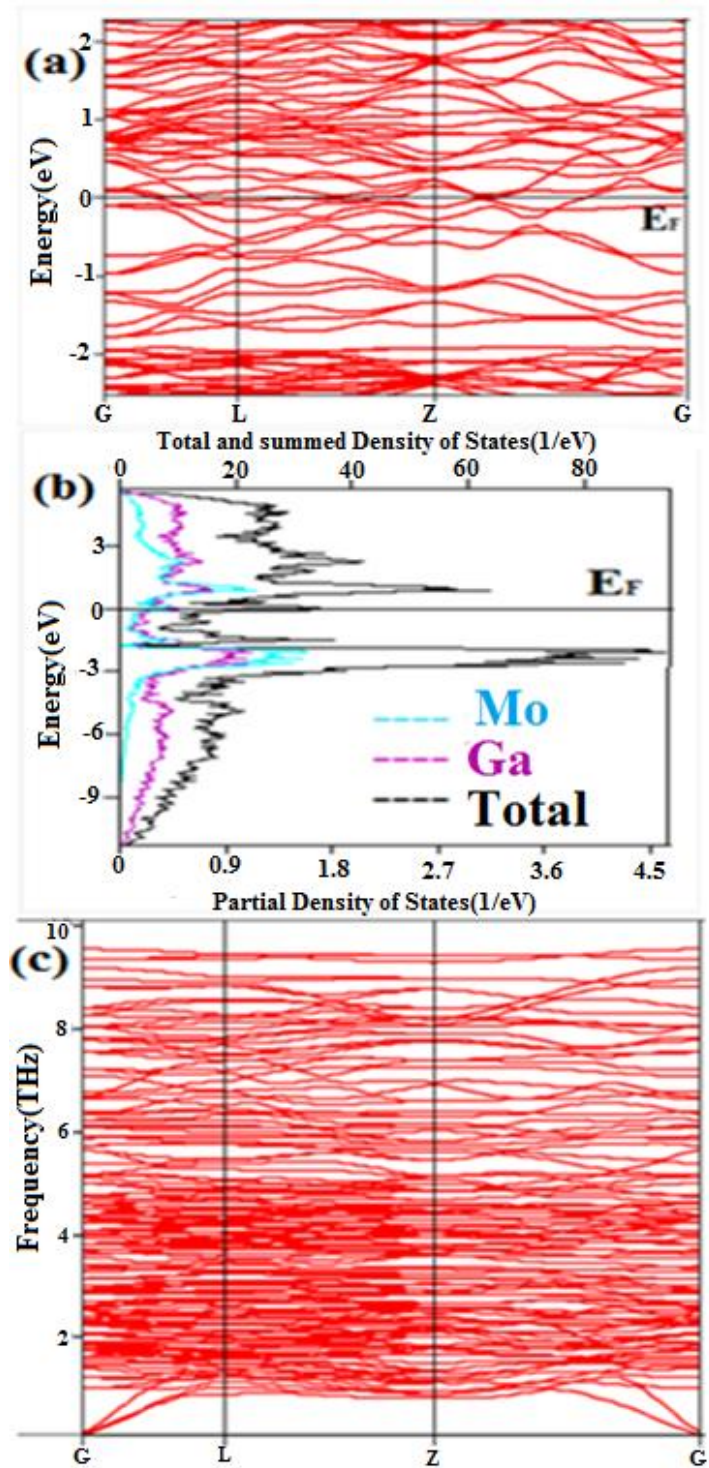

**Figure 3**

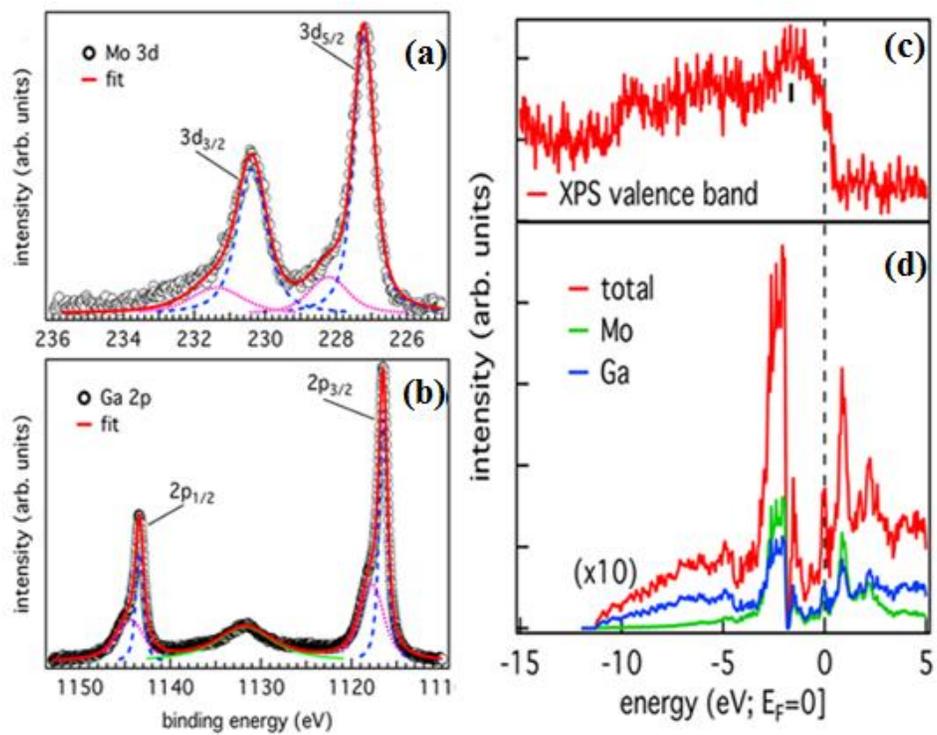

**Figure 4**

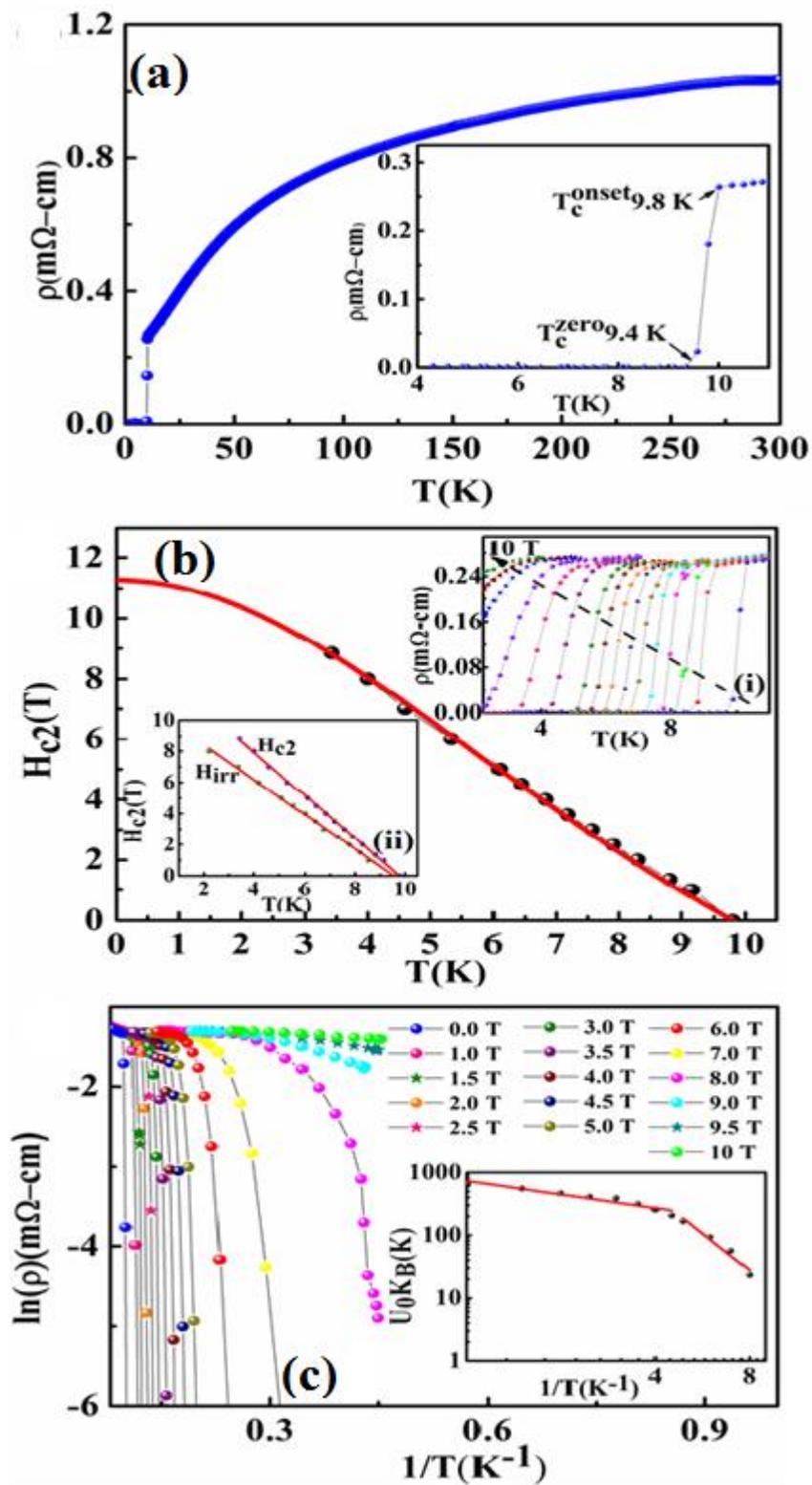

**Figure 5**

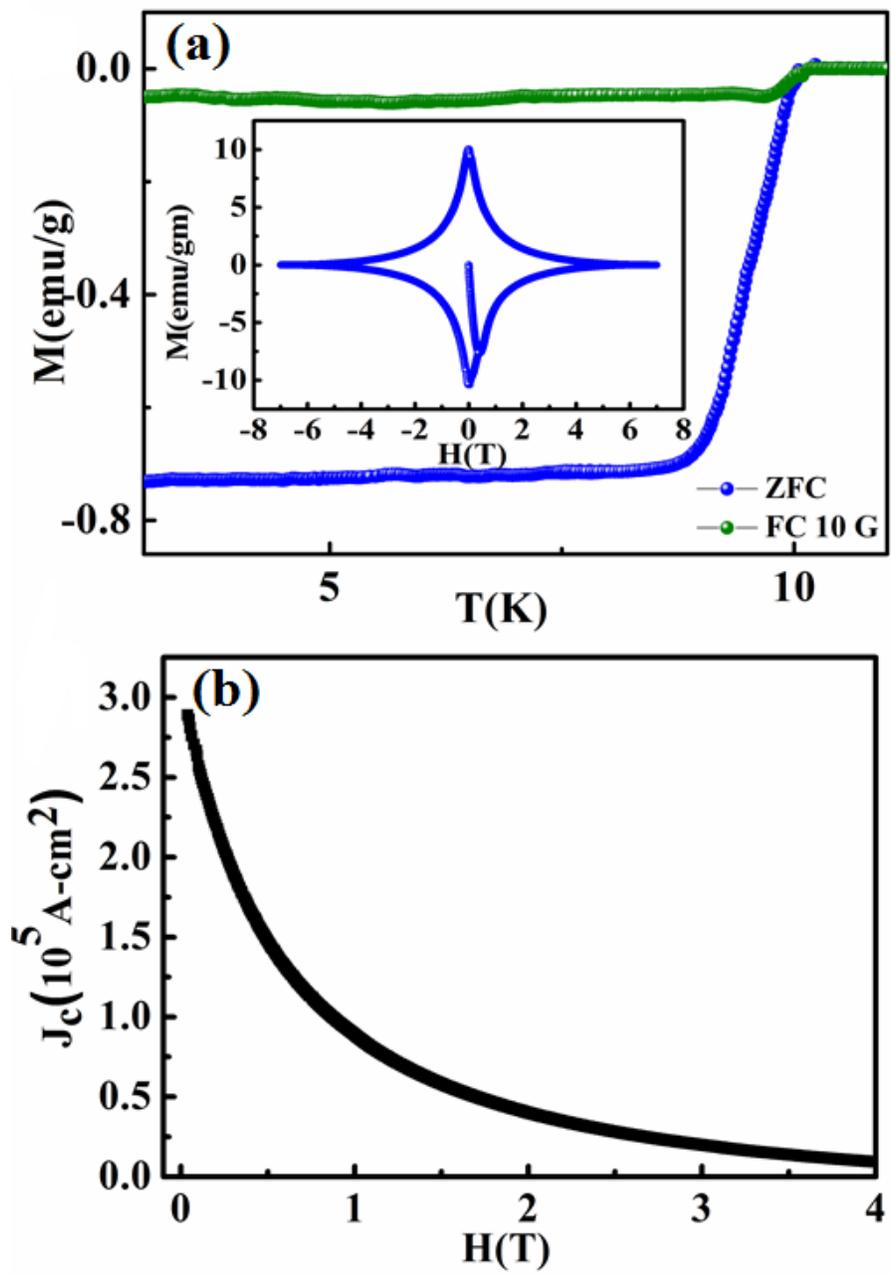

**Figure 6**

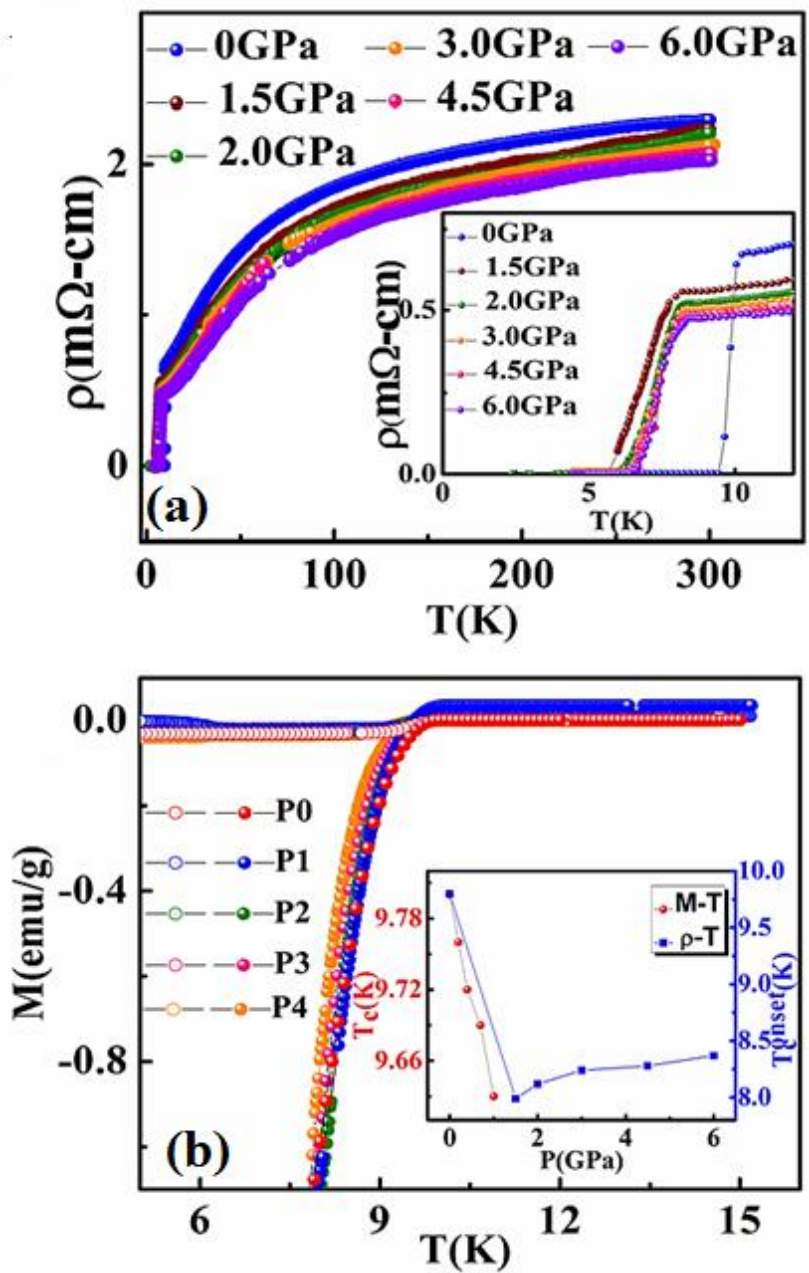

**Figure 7**

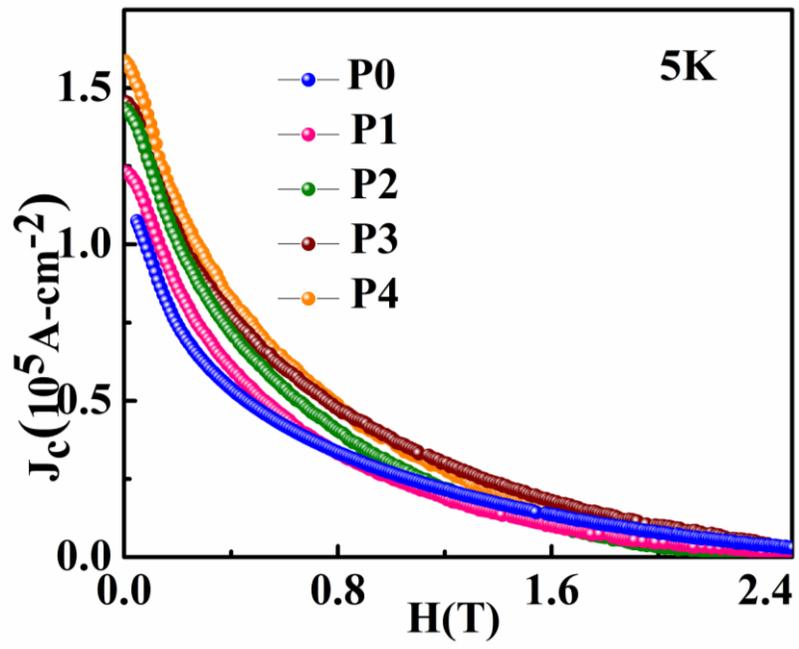

**Figure 8**

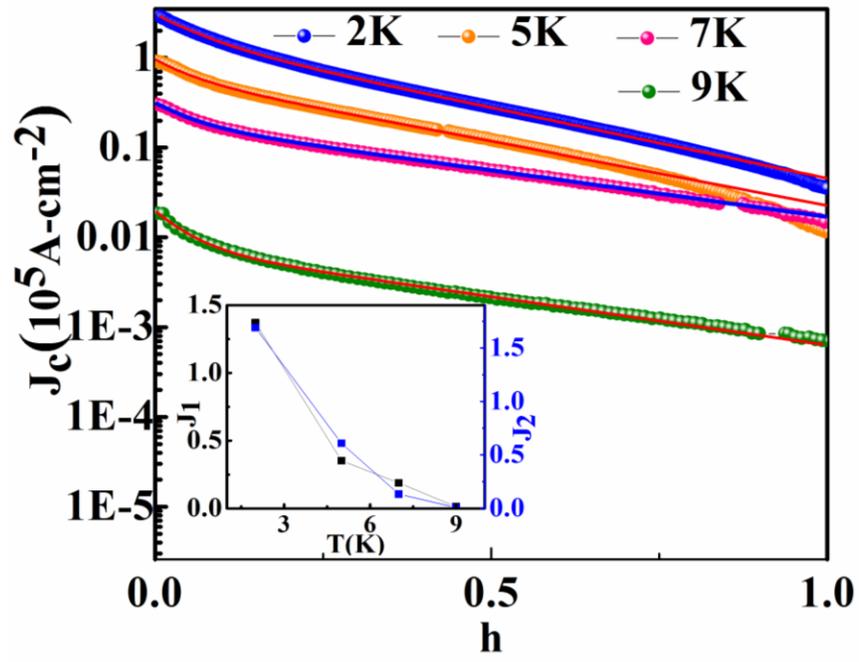